\documentclass[sigplan, nonacm]{acmart}

\acmYear{2019}
\acmISBN{} 
\acmDOI{} 
\startPage{1}

\setcopyright{none}

\usepackage{booktabs}   
\usepackage{subcaption} 

\usepackage{omega}
\usepackage[scaled=0.75]{beramono}
\usepackage{xspace}

\usepackage{wrapfig}
\usepackage{amsfonts}
\usepackage{amsmath}
\usepackage{amssymb}
\usepackage{stmaryrd}
\usepackage{graphicx}
\usepackage{fancyvrb}
\usepackage{amsthm}
\usepackage{mathtools}
\usepackage{bm}
\usepackage{array}
\usepackage{enumitem}
\usepackage{proof}

\theoremstyle{definition} 
\newtheorem{definition}{Definition}

\newtheorem{theorem}{Theorem}
\graphicspath{{figures/}}

\newcommand{\inlinecode}{\texttt}

\DeclareMathOperator{\hire}{offer}

\DeclareMathOperator{\Real}{Real}
\DeclareMathOperator{\Int}{Int}
\DeclareMathOperator{\Bool}{Bool}

\DeclareMathOperator{\rcd}{rcd}

\DeclareMathOperator{\ciid}{ciid}

\DeclareMathOperator{\rand}{rand}
\DeclareMathOperator{\true}{true}
\DeclareMathOperator{\false}{false}

\DeclareMathOperator{\uuid}{uid}

\DeclareMathOperator{\SU}{\mathcal{S}^U}

\DeclareMathOperator{\bern}{Bernoulli}
\DeclareMathOperator{\unif}{Uniform}

\DeclareMathOperator{\stdunif}{unif}

\DeclareMathOperator{\normal}{\mathcal{N}}
\DeclareMathOperator{\ee}{\mathbb{E}}
\DeclareMathOperator{\prob}{\mathbb{P}}
\DeclareMathOperator{\rv}{rv}
\DeclareMathOperator{\RV}{RV}

\DeclareMathOperator{\dist}{dist}

\DeclareMathOperator{\dq}{:=}

\newcommand{\RVT}[1]{\RV #1}

\newcommand{\indep}{\perp \!\!\!\perp}

\newcommand{\omegalang}{\textsc{Omega}}

\newcommand{\lifted}[1] {\tilde{#1}}

\newcommand{\lmean}[1] {\lifted{\ee}{(#1)}}
\newcommand{\conds}[2] {#1  \mid #2}
\newcommand{\cnd}[2] {#1  \mid #2}
\newcommand{\rcdxy}[2] {#1 \parallel #2}

\newcommand{\ciidset}[2]{\left[#1\right]_{\indep \mid #2}}

\usepackage[colorinlistoftodos,prependcaption,textsize=tiny]{todonotes}
\usepackage{xargs} 

\newcommandx{\unsure}[2][1=]{\todo[linecolor=red,backgroundcolor=red!25,bordercolor=red,#1]{#2}}
\newcommandx{\change}[2][1=]{\todo[linecolor=blue,backgroundcolor=blue!25,bordercolor=blue,#1]{#2}}
\newcommandx{\info}[2][1=]{\todo[linecolor=OliveGreen,backgroundcolor=OliveGreen!25,bordercolor=OliveGreen,#1]{#2}}
\newcommandx{\improvement}[2][1=]{\todo[linecolor=Plum,backgroundcolor=Plum!25,bordercolor=Plum,#1]{#2}}
\newcommandx{\wtp}[2][1=]{\todo[linecolor=orange,backgroundcolor=orange!25,bordercolor=orange,#1]{#2}}
\newcommandx{\thiswillnotshow}[2][1=]{\todo[disable,#1]{#2}}

\usepackage{algorithm}
\usepackage{algorithmic}
                        
\begin{document}

\title[The Random Conditional Distribution]{The Random Conditional Distribution}        
\subtitle{For Higher-Order Probabilistic Inference}                     


\author{Zenna Tavares}
\affiliation{
  \department{BCS, CSAIL}              
  \institution{MIT}            
}

\author{Xin Zhang}
\affiliation{
  \department{CSAIL}              
  \institution{MIT}            
}

\author{Edgar Minaysan}
\affiliation{
  \institution{Princeton}            
}

\author{Javier Burroni}
\affiliation{
  \institution{UMass Amherst}            
}

\author{Rajesh Ranganath}
\affiliation{
  \department{Courant Institute}              
  \institution{NYU}            
}

\author{Armando Solar Lezama}
\affiliation{
  \department{CSAIL}              
  \institution{MIT}            
}

\begin{abstract}
The need to condition distributional properties such as expectation, variance, and entropy arises in algorithmic fairness, model simplification, robustness and many other areas.
At face value however, distributional properties are not random variables, and hence conditioning them is a semantic error and type error in probabilistic programming languages.
On the other hand, distributional properties are contingent on other variables in the model, change in value when we observe more information, and hence in a precise sense are random variables too.
In order to capture the uncertain over distributional properties, we introduce a probability construct -- the random conditional distribution -- and incorporate it into a probabilistic programming language \omegalang{}.
A random conditional distribution is a higher-order random variable whose realizations are themselves conditional random variables.
In \omegalang{} we extend distributional properties of random variables to random conditional distributions, such that for example while the expectation a real valued random variable is a real value, the expectation of a random conditional distribution is a distribution over expectations.
As a consequence, it requires minimal syntax to encode inference problems over distributional properties, which so far have evaded treatment within probabilistic programming systems and probabilistic modeling in general.
We demonstrate our approach case studies in algorithmic fairness and robustness.

\end{abstract}




\maketitle

\section{Introduction}

Probabilistic programming languages encode probabilistic models as programs. They are the most expressive among a long lineage of formalisms including Bayesian networks, factor graphs, and systems of statistical equations.
A formalism is more expressive if it can represent a larger class of models or permits a wider range of inference queries.
A class of more recent languages including Church \cite{goodman2008church}, Anglican \cite{wood2014new} and Venture \cite{mansinghka2014venture} have tended towards one extreme of the spectrum, allowing recursion, control flow, and unbounded (or even random) numbers of discrete, continuous, or arbitrarily-typed random variables.
These languages extend Turing complete deterministic languages to attain a notion of probabilistic universality: the ability to express any computable probability distribution.

There is a a blind spot -- in terms of expressiveness -- in even universal probabilistic programming languages.
Existing probabilistic languages provide mechanisms to define random variables, transform them and condition them.
Many languages also provide mechanisms to compute \emph{distributional properties} such as expectation, variance, and entropy, or approximate them from samples.
However, they lack any automated or composable mechanisms to capture the uncertainty over such distributional properties, which is a major limitation.

Distributional properties are fixed (often real) values, but in a sense they are random variables too.
For example, rainfall depends on temperature, the season, the presence of clouds, and so on.
Probabilistic models are routinely used to capture uncertainty over these factors and their complex interactions.
With respect to a model, expected rainfall is a real value, but it changes if we obtain new information.
For example it rises if we observe clouds and falls to zero if we observe their absence.
These two expectations become a random variable over expectations -- a \emph{conditional expectation} -- when we take into account the probabilities of the presence or absence of clouds.
Moreover, for each random variable in the model there is a corresponding conditional expectation.
For instance, with respect to the season, conditional expected rainfall is a random variable over four expectations, one for each season; with respect to temperature it is a continuous distribution.
These conditional expectations capture the uncertainty over expected rainfall that results from other variables in the model, whereas the unconditional expected rainfall averages all the uncertainty away.

The ability to capture uncertainty over distributional properties is useful for the same reason that distributions are preferable to point estimates in general: they possess more information.
In addition, conditioning distributional properties opens up even more applications.
For example, algorithmic fairness has received interest recently due to the expansion of machine learning into sensitive areas such as insurance, mortgages and employment.
One probabilistic formulation is \emph{equality of opportunity} \cite{equaoppo}, which states that the probability a qualified member in a minority group $v_m$ receives a positive classification, is not far from the probability a qualified member in a majority group $v_n$ receives a positive classification.
These probabilities are distributional properties (probability is a special case of expectation).
Capturing the uncertainty over them (randomizing them) allows us to condition a probabilistic classifier to satisfy equality of opportunity, rather than just verify whether it does or not.

Conditioning distributional properties is often computationally very challenging, but as inference procedures develop a number of problems could be unified under the same framework.
For instance, classifications from deep neural networks have been shown to be vulnerable to adversarial attacks \cite{szegedy2013intriguing}.
This could be mitigated by simply conditioning on the property that small random perturbations cannot dramatically change the classification distribution
Another use case is to construct priors which adhere to distributional constraints.
For example if we only have weak prior knowledge about a variable, such as its mean and variance, we can use a flexible nonparametric distribution family such as a Bayesian neural network (a neural network with a distribution over the weight parameters) but fix its mean and variance using random conditional distributions.
Distributional properties may be functions of more than one random variable, for instance the Kullback-Leibler divergence and Wasserstein distance are functions of two random variables.
Bounding divergences using random conditional distributions could be used for model simplification: to condition on a simple distribution being not far in divergence from a much more complex one.



As the primary contribution of this paper we introduce the \emph{random conditional distribution} to randomize distributional properties.
Given two random variables $X$ and $\Theta$, the random conditional distribution of $X$ given $\Theta$ -- which we denote $\rcdxy{X}{\Theta}$ -- is a a \emph{random distribution}: a random variable whose realizations are themselves random variables.
In particular, each realization of $\rcdxy{X}{\Theta}$ is the random variable $X$ conditioned on $\Theta = \theta$ where $\theta \sim \Theta$ is a realization of $\Theta$.

Intuitively, the random conditional distribution decomposes a probabilistic model into a distribution over probabilistic models.
For example, if $\Theta=\bern(0.4)$ and $X = \normal(\Theta, 1)$, then $\rcdxy{X}{\Theta}$ is a random conditional distribution comprised of two normally distributed random variables $\conds{\normal(\Theta, 1)}{\Theta = 0}$ and $\conds{\normal(\Theta, 1)}{\Theta = 1}$.
The probabilities that $\rcdxy{X}{\Theta}$ takes these different outcomes is determined by the prior probabilities of the different outcomes of $\Theta$: $0.4$ and $0.6$ respectively.
We extend distributional properties to random conditional distributions such that while the expectation $\ee(X)$ is a real value, $\ee(\rcdxy{X}{\Theta})$ is a distribution over real values (expectations).
In particular, it is the distribution from taking the expectation of each of the two normally distributed random variables in $\rcdxy{X}{\Theta}$: 0 and 1 respectively.

The extension of distributional properties to random conditional distributions allows us to randomize distributional properties with minimal syntax.
Returning to the fairness example, let $\hire(v_m, \Theta)$ be a distribution over whether a candidate drawn from the distribution $v_m$ will receive a job offer, where $\Theta$ is a distribution over weights of a probabilistic classifier.
The classifier satisfies equality of opportunity if the ratio of probabilities of receiving an offer between groups is bounded by $\delta$:
\begin{equation}\label{fair}
\frac{\prob(\hire(v_n, \Theta))}{\prob(\hire(v_m, \Theta))} < \delta
\end{equation}
To construct a fair classifier rather than determine whether a classifier is fair,  we want to condition on it being fair, i.e., find the conditional distribution over $\Theta$ given that Expression (\ref{fair}) holds.
Again, this is problematic because Expression (\ref{fair}) is not a random variable and hence cannot be conditioned, despite being composed of random variable parts.
$\Theta$, $v_m$, and hence $\hire(v_m, \Theta)$ are all random variables, but $\prob(\hire(v_m, \Theta))$ is a real value (the probability someone drawn from $v_m$ receives a job offer), and Expression (\ref{fair}) is a Boolean -- the classifier is either fair or not.
In contrast, $\prob(\rcdxy{\hire(v_m, \Theta)}{\Theta})$ is not a probability but a distribution over probabilities that $v_m$ receives a job offer.
Similarly, to turn expression 1 from a Boolean into a Boolean-valued random variable that can be conditioned requires only a small change:
\begin{equation}\label{eq:fairrcd}
\frac{\rcdxy{\prob(\hire(v_n, \Theta)}{\Theta})}{\rcdxy{\prob(\hire(v_m, \Theta)}{\Theta})} < \delta 
\end{equation}

While random conditional distributions can greatly improve the expressiveness of probabilistic programming, do we need new probabilistic languages, or can random conditional distributions be implemented in existing ones?
We find that although one can explicitly define random conditional distributions for specific situations in existing languages, it is not possible to define a generic function $\parallel$ which automatically induces random conditional distributions from the original model.
Based on this, we have designed a probabilistic programming language \omegalang{} for distributional inference using random conditional distributions.

In summary, we address the neglected problem of distributional inference: the randomizing and conditioning of distributional properties.
Our method depends on the random conditional distribution, a higher-order probability construct.  These concepts are synthesized in a probabilistic programming language \omegalang{}, which has an operator to construct random conditional distributions as a primitive construct.  In more detail:
\begin{enumerate}[topsep=0pt,itemsep=-1ex,partopsep=1ex,parsep=1ex]
	\item We formalize random conditional distributions (section \ref{rcd}) within measure theoretic probability.
	\item We present the syntax and semantics of a probabilistic programming language \omegalang{} for distributional inference (section \ref{olang}) using random conditional distributions.
	\item Finally, we demonstrate \omegalang{} using several representative benchmarks including two case studies, where it is applied to infer a fair classifier and where it is applied to infer a classifier that is robust to adversarial inputs (section \ref{exp}).
\end{enumerate}

\section{Background on Uncertain Probabilities}\label{higher}


Probabilities over probabilities, often called higher-order probabilities \cite{gaifman1988theory}, subsumed most of the inquiry into the interpretation of uncertain distributional properties. 
Interest in higher-order probabilities was motivated by an apparent failure of standard probability theory to distinguish between uncertainty and ignorance.
For instance, a weather forecaster may project a 30\% chance of of rain, which is straightforwardly captured as a first-order probabilistic statement: $\prob(rain) = 0.3$.
However, she may feel only 70\% confident in that assessment.
How to both express and interpret degrees of confidence in probabilistic statements within probability theory has led to much debate and disagreement.
  
Pearl \cite{pearl1987we} summarized several difficulties with higher-order probabilities.
Traditional probability theory requires that probabilities are assigned to factual events that in principle could be verified by empirical tests.
Under this principle it is problematic to assign probabilities to probability distributions themselves, because the truthfulness of probabilistic statements cannot be determined empirically.
The statement \linebreak$\prob(rain) = 0.3$ could be verified, albeit with much difficulty, given sufficient knowledge of the physical systems which govern the weather.
In contrast, the second order statement describing the weather forecaster's confidence, denoted by $\prob\left[\prob(rain) = 0.3\right] = 0.7$ (assuming momentarily this is syntactically valid), appears much less amenable to scrutiny by empirical test.
What do probabilities of probabilities mean then, if their truthfulness cannot be determined empirically even in principle?

Several probabilistic \cite{domotor1981higher, gaifman1988theory} and non-probabilistic \cite{dempster2008upper, shafer1976mathematical} formalisms have been developed to unify uncertainty and confidence.
However, Pearl \cite{pearl1987we} and Kyburg \cite{kyburg2013higher} argued that standard first-order probability suffices; neither second-order nor higher-order extensions to probability theory were necessary.
In particular, probabilistic statements of the form $\prob(A) = p$ are themselves empirical events.
To say such an event occurred means roughly to say one mentally computed that the probability of $A$ is $p$.
Hence events of this kind are subjective, and although not amenable to public scrutiny, are of no lesser stature than any other event.

The question of what renders a statement such as $\prob(A) = p$ an unknown, random event, rather than a fixed outcome of a deterministic procedure, remains.
A resolution first presented by de Finetti \cite{de1977probabilities} suggested that $\prob(A) = p$ is a random event whenever the assessment of $\prob(A)$ depends substantially on other events in the system.
For instance the assessment $\prob(rain) = 0.3$ is uncertain because it depends substantially on $\prob(clouds)$; the occurrence or non-occurrence of clouds would dramatically change our confidence in $\prob(rain) = 0.3$.
Pearl \cite{pearl1987we} formalized this notion of dependence within the framework of causal probabilistic models, asserting that an event $A$ substantially depends on $B$ if $B$ is a cause of $A$ with respect to a causal model.
Crucially, Pearl demonstrated that the causal model provides both necessary and sufficient information for computing both $\prob(A) = p$ and $\prob\left[\prob(A) = p\right]$. In other words, higher-order probabilities exist, but they are derived entirely from the original model.

Modern measure theoretic probability uniformly accommodates first and higher-order statements.
In particular, since the probability of an event is the expectation of the random variable that indicates it, higher-order probabilities are conditional expectations.
That is, uncertainty over a probability $\mathbb{P}(A)$ is captured by the conditional expectation $\ee(\mathbf{1}_A \mid C)$ with respect to some contingency set $C$, where $\mathbf{1}_A(\omega)$ is 1 if $\omega \in A$ and 0 otherwise.
Still, several issues remain.
Namely, how to (i) generalize from higher-order probabilities to all distributional properties (ii) operationalize them within probabilistic programming languages so that they can be used in practice.
\section{Background}\label{basics}

In this section we introduce the foundations for our approach, which is largely measure-theoretic probability \citep{ccinlar2011probability}.

\paragraph{Random Variables.} Probability models lie on top of probability
spaces. A probability space is a measure space $(\Omega, \mathcal{H}, \mathcal{P})$,
where $\Omega$ is the sample space, $\mathcal{H}$ is a $\sigma$-algebra over subsets of $\Omega$, and $\mathcal{P}$ is the probability measure ($\mathcal{P}(\Omega) = 1$). Random variables are
functions from the sample space $\Omega$ to a realization space $\tau$.
A model is a collection of random variables along with a probability space. 
A concrete example of a probability space takes $\Omega$ to be hypercube, with $\mathcal{P}$ being uniform a uniform measure over that hypercube.
A normally distributed random variable maps from $\Omega \to \mathbb{R}$.
Since the underlying probability space is uniform, this function is the inverse cumulative distribution function of the normal.

\paragraph{Random Variable Algebra} It is convenient to treat random variables (which are functions) as if they were values from their realization space.
For example, if $X$ is a random variable, then $X + X$ and $X / 3$ are also random variables.
The semantics of this syntax defines operations on random variables pointwise.
For example: $(X+X)(\omega) = X(\omega) + X(\omega)$.
More generally, let $X:\Omega \to \tau_1$ and $f:\tau_1 \to \tau_2$ be a function, then $Y = f(X)$ is a random variable defined as:
\begin{equation}\label{eq:pointwise}
Y(\omega) = f(X(\omega))
\end{equation}

\paragraph{Conditioning}


Conditioning restricts a model to be consistent with a predicate.
It can be operationalized as an operator $\mid$ that concentrates the probability space of a random variable $X:\Omega \to \tau$ to an event $A$ indicated by a predicate $Y$, i.e.: $A = \{ \omega \mid Y(\omega) = 1\}$.
That is, $\conds{X}{Y}: \Omega \to \tau$ is functionally identical to $X$ -- $(\conds{X}{Y})(\omega) = X(\omega)$ -- but defined on the probability space 
$(\Omega \cap A, 
\{A \cap B \mid B \in \mathcal{H} \}, \mathcal{P} / \mathcal{P}(A))$, the concentration of $\mathcal{S}$ onto $A$.

The general construction of new models might require conditioning
on sets of measure zero. This process can be made rigorous
via disintegration \citep{chang1997conditioning}, which can
be thought of as the reversal of building joint distributions through
product measure constructions.


\paragraph{Distributional Properties and Operators}
Distributional properties such as expectation, variance and entropy summarize aspects of a random variable.
We use the term distributional operator for the higher-order function that maps random variables to the corresponding distributional property.
For example the expectation distributional operator $\ee : (\Omega \to \mathbb{R}) \to \mathbb{R}$ maps a random variable $X$ to its expectation $\ee(X)$.  It is rigorously defined in terms of Lebesgue integration over the sample space $\Omega$ with respect to the probability measure $\mathcal{P}$:
$$
\ee(X) = \int _{\Omega }X(\omega )\,d \mathcal{P} (\omega )
$$

Distributional properties are in many cases intractable to compute exactly, but can be approximated from samples.

\section{The Random Conditional Distribution}\label{rcd}
Random conditional distributions provide a mechanism to condition distributional properties.
Given two random variables $X$ and $\Theta$, the random conditional distribution of $X$ given $\Theta$ -- which we denote $\rcdxy{X}{\Theta}$ -- is a a \emph{random distribution}: a random variable whose realizations are themselves random variables.
In particular, each realization of $\rcdxy{X}{\Theta}$ is the random variable $X$ conditioned on $\Theta = \theta$ where $\theta \sim \Theta$ is a realization of $\Theta$:

\begin{definition}The random conditional distribution (rcd) of a random variable $X: \Omega \to \tau_1$ given $\Theta: \Omega \to \tau_2$ is a random variable $\rcdxy{X}{\Theta}: \Omega \to (\Omega \to \tau_1)$, defined as:
\begin{equation}\label{eq:rcd}
(\rcdxy{X}{\Theta})(\omega) \dq \conds{X}{\Theta = \Theta(\omega)}
\end{equation}

Intuitively, the random conditional distribution decomposes a probabilistic model into a distribution over probabilistic models.
For example, if $\Theta=\bern(0.4)$ and $X = \normal(\Theta, 1)$, then $\rcdxy{X}{\Theta}$ is a random conditional distribution comprised of two normally distributed random variables $\conds{\normal(\Theta, 1)}{\Theta = 0}$ and $\conds{\normal(\Theta, 1)}{\Theta = 1}$.
The probabilities that $\rcdxy{X}{\Theta}$ takes these different outcomes is determined by the prior probabilities of the different outcomes of $\Theta$: $0.4$ and $0.6$ respectively.
\end{definition}

A consequence of random variable algebra (section \ref{basics}) is that applying distributional operators to random condition distributions yields random distributional properties.
Continuing the example from above, $\ee(\rcdxy{X}{\Theta})$ is a random variable over expectations taking values 0 and 1 with probabilities 0.4 and 0.6 respectively.
$\ee(\rcdxy{X}{\Theta})$ is a conditional expectation, denoted $\ee(X \mid \Theta)$ in standard mathematical notation.

\begin{theorem}\label{thm:rcdeverywhere}
$\ee(\rcdxy{X}{\Theta})$ is the conditional expectation of $X$ with respect to $\Theta$ defined as $\ee(X \mid \sigma(\Theta))(\omega) = \ee(X \mid \Theta = \Theta(\omega))$.
\end{theorem}
\begin{proof}
For clarity we distinguish expectation defined on real valued random variables $\ee$ from expectation defined on real valued random distributions $\tilde{\ee}$:
\begin{equation*}
\lmean{\rcdxy{X}{\Theta}} = \lmean{\lambda \omega. (\conds{X}{\Theta = \Theta(\omega)})} 
                    = \lambda \omega . \ee(\conds{X}{\Theta = \Theta(\omega)})
\end{equation*}
\end{proof}

The expectation of a random conditional distribution is a conditional expectation, but the same mechanism works for any distributional property.
If $\mathbb{O}$ is a distributional operator defined on $\tau$ valued random variables, then it extends pointwise to random conditional distributions:
$$
\mathbb{O}(\rcdxy{X}{\Theta})(\omega) = \mathbb{O}((\rcdxy{X}{\Theta})(\omega)) = \mathbb{O}(\conds{X}{\Theta = \Theta(\omega)})
$$

$\mathbb{O}$ could be expectation, variance, entropy, information, or support.
$\mathbb{O}$ could also map from more than one random variable, such as KL-Divergence or mutual information.

It is possible that uncertainty over a distributional property is captured by a variable already in the model, which is simpler than our construction using random conditional distributions.
For example, if $\Theta = \unif(0,1)$ and $X = \bern(\Theta)$, then $\Theta$ is equal to the conditional expectation $\ee(\cnd{X}{\Theta})$, and therefore equal to $\ee(\rcdxy{X}{\Theta})$.
This is a consequence of the fact that the weight parameter of a Bernoulli distribution is also its expectation.

However, this is a special case; for most distributional properties, most models do not possess a corresponding variable.
For example the variance of $X$ is not a variable in example model above, and if we substitute the $\bern(\Theta)$ with $\textrm{Beta}(\Theta, 1)$, neither is the expectation.
In some cases a distributional property that is not in the model can be derived as a transformation of existing variables.
Continuing the example, if $X = \textrm{Beta}(\Theta, 1)$, then its expectation is the random variable $\Theta / (\Theta + 1)$.
This however, is also not always possible.  In many models there exists no closed form expression for a given distributional property.
For example if we instead have $X = \textrm{Gamma}(\Theta, 1)$, there is no closed form expression from the median.

Even when it is possible, a further limitation of this approach is that it lacks the flexibility afforded by the second argument of rcd.
The uncertainty in $\mathbb{O}(\rcdxy{X}{\Theta})$ is a consequence of uncertainty in $\Theta$.
Changing $\Theta$ reveals different perspectives and granularities of the uncertainty.  For example, if deciding when to take a mountaineering trip one might be interested in $\ee(\rcdxy{rain}{season})$, but if deciding how tall of a mountain to tackle then $\ee(\rcdxy{rain}{altitude})$ may be more informative.
Moreover $\Theta$ can be crucial when conditioning distributional properties.
In the equality of opportunity example from the introduction, if we instead find the conditional distribution over $\Theta$ given that
\begin{equation}\label{eq:fairrcdbad}
  \frac{\rcdxy{\prob(\hire(v_n, \Theta)}{(\Theta, v_n, v_m)})}{\rcdxy{\prob(\hire(v_m, \Theta)}{(\Theta, v_n, v_m)})} < \delta 
\end{equation}  
holds, we would in effect eliminate portions of the population.
This is clearly not what is meant by fairness.

\section{The \omegalang{} Programming Language}\label{olang}

This section describes the core language of \omegalang{}:
a functional language augmented with a small number of probability constructs.
Most of the probabilistic constructs -- random variables,  probability spaces and conditional independence -- correspond directly to concepts in measure-theoretic probability.
Figure \ref{syntax} shows the abstract syntax of \omegalang{}.

\subsection{Types}\label{SYNTAXTYPES}

\begin{figure}
	{\small
	\begin{align*}
	\text{ type }&&   \tau ::= &\Int \mid \Bool \mid \Real \mid \Omega \mid \RV \tau \mid \tau_1 \to \tau_2 \mid (\tau_1, \tau_2) \\
	\text { term }&&  t ::= &n \mid b \mid r \mid \bot \mid x \mid  \text{ if } t_1 \text{ then } t_2 \text{ else } t_3 \mid t_1(t_2) \mid \\
	&& &(t_1, t_2) \mid \text{let } x = t_1 \text{ in } t_2  \mid \lambda x : \tau. t \mid \\
    &&  &\rv(t) \mid \ciid(t) \mid \uuid \mid (\cnd{t_1}{t_2})  \mid \rcdxy{t_1}{t_2} \mid \dist(t) \\
	\text{ query }&& &\rand(t)
	\end{align*}
}
\vspace{-0.3in}
	\caption{Abstract Syntax for \omegalang{}}
	\label{syntax}
\end{figure}

The type system is composed of primitives $\Int$, $\Bool$, and $\Real$ with their standard interpretation.
There are two primitive probabilistic types:
$\Omega$ the sample space type, and $\RV \tau$ the $\tau$-valued random variable type.
Function types are expressed with $\tau \to \tau$. \omegalang{} is curried, and hence multivariate functions have nested function types of the form $\tau \to \cdots \to \tau$.
Tuple types are denoted with $(t_1, t_2)$.
Every type is lifted to include $\bot$, the undefined value, which plays an important role in conditioning.
Typing rules on probabilistic terms are given in figure \ref{typing}.

\subsection{Syntax}\label{SYNTAX}

Standard terms have their standard interpretation.
$n$ represents integer numbers, $b$ are Boolean values in $\{\true, \false\}$, and $r$ are real numbers.
$x$ represents a variable in a set of variable names $\{\omega, x, y, z, \dots\}$.
$\bot$ represents the undefined value.
$\lambda x : \tau . t$ is a lambda abstraction; $t_1(t_2)$ is function application;
conditionals are expressed with $\text{ if } t_1 \text{ then } t_2 \text{ else } t_3$; variable bindings are defined with let.

\omegalang{} has several probabilistic terms.
If $t$ is a function of type $\Omega \to \tau$, $\rv(t)$ is a random variable.
$\ciid(t)$ creates a new random variable that is distributed identically to $t$, but is conditionally independent given parent random variables it depends on.
$\uuid$ is a macro (i.e. resolved syntactically) that evaluates  unique integer.
It is is used to extract a unique element of the sample space.
$\cnd{t_1}{t_2}$ constructs a conditional random variable equivalent to $t_1$ but defined on a sample space restricted by $t_2$.
The random conditional distribution of $t_1$ given $t_2$ is denoted by $\rcdxy{t_1}{t_2}$.
The query term $\rand(t)$ draws a sample from a random variable $t$.

Operators $\mid$ and $\parallel$ map from and to arbitrarily-valued random variables, and  hence are higher-order functions with polymorphic types.
For simplicity of presentation \omegalang{} lacks user-defined polymorphic types, but we implement these built-in polymorphic functions (figure \ref{macros}).

Operator $\dist$ represents broadly functions that evaluate distributional properties of random variables.
Its type is $\RV T\linebreak \to \Real$ where $T$ is a primitive type.
We rely on the underlying inference engine to provide these functions and treat $\dist$ as a blackbox
in this section.
In particular, we focus on functions that are deterministic and can be approximated using sampling,
such as expectation, variance, and KL divergence.

\begin{figure}[t]
\begin{align*}
\textbf{Macros:} \\
\uuid &= \text{ integer unique across program }\\
\textbf{Functions:} \\
\conds{x}{y} &= \lambda x:\RV \tau, \lambda x:\RV \Bool . \\
& rv(\lambda \omega:\Omega.\text{ if } y(\omega) \text{ then } x(\omega) \text{ else } \bot)\\
\rcdxy{x}{\theta} &= \lambda:\RV \tau_1, \theta:\RV \tau_2 . \lambda \omega : \Omega . \conds{x}{\theta = \theta(\omega)}
\end{align*}
\caption{Built-in functions and macros in \omegalang{}}
\label{macros}
\end{figure}

\begin{figure}[h]
	{\small
\begin{gather*}
    \infer[rv]{\Gamma \vdash \textrm{rv}(t_1) : \RVT{\tau}}
    {\Gamma \vdash t_1 : \Omega \to \tau}
    \quad \quad \quad
    \infer[ciid]{\Gamma \vdash \ciid(t_1) : \RVT{\tau}}
    {\Gamma \vdash t_1 : \RVT{\tau}}
    \\
    \infer[cond]{\Gamma \vdash \cnd{t_1}{t_2} : \RVT{\tau}}
    {\Gamma \vdash t_1 : \RVT{\tau} & \Gamma \vdash t_2 : \RVT{\Bool}}
    \\
    \infer[rcd]{\Gamma \vdash \rcdxy{t_1}{t_2} : \RVT{\RVT{\tau_1}}}
    {\Gamma \vdash t_1 : \RVT{\tau_1} & \Gamma \vdash t_2 : \RVT{\tau_2}}
    \\
    \infer[rvapp]{\Gamma \vdash  (t_1 \; t_2) : \RVT{\tau_2}}
    {\Gamma \vdash t_1 : \tau_1 \to \tau_2 & \Gamma \vdash t_2 : \RVT{\tau_1}}
    \\
    \infer[rand]{\Gamma \vdash \rand(t) : \tau}
    {\Gamma \vdash t : \RVT{\tau}}
    \\
\end{gather*}    }
\vspace{-0.4in}
    \caption{Typing rules of \omegalang{}}
    \label{typing}
\end{figure}
\subsection{Denotational Semantics}\label{densemsec}
\newcommand{\sem}[1]{\llbracket #1 \rrbracket}
\newcommand{\den}[1]{\left[#1\right]}

Here we define a denotational semantics of \omegalang{} which is shown in figure \ref{densemantics}.
In the denotational semantics, standard terms are standard and omitted.
The denotation $\sem{t}$ of a term $t$ is a value in a semantic domain corresponding to an \omegalang{} type, such as a Boolean, real number, or random variable.

\subsubsection*{Semantic Domains}

For the purpose of the denotational semantics, we assume an unconditioned probability space $\SU = (\Omega, \mathcal{H}, \mathcal{P})$, $\Omega$ is a $d$-dimensional unit hypercube $[0, 1]^d$:
$$
\Omega = \Omega_1 \times \Omega_2 \times \cdots \times \Omega_d \text{ where } \Omega_i = [0, 1]
$$

If $\omega \in \Omega$, then $\omega_i$ denotes the $i$th component and is a real value in $[0, 1]$.
$\mathcal{P}$ is any probability measure such that the set of random variables $\{\lambda \omega :\Omega . \omega_i \mid \forall i\}$ are mutually independent and uniformly distributed on $[0, 1]$.

$\SU$ is natural choice of a probability space because any univariate probability distributions can be constructed by transformation of a uniform distribution over the unit interval, and arbitrary multivariate random variables can be defined by transformations of the unit hypercube.

Different random variables may be conditioned on different events, and hence defined on different probability spaces.
Specifically, different random variables may be defined on different concentrations of the unrestricted space $\SU$.
If $A$ is some event of positive measure, $\mathcal{S}' = \SU \cap A$ is the concentration of $\SU$ onto $A$:
$$
\SU \cap A = (\Omega \cap A, 
\{A \cap B \mid B \in \mathcal{H} \}, \mathcal{P} / \mathcal{P}(A))
$$

Random variables exist in a parent-child relation.
This relation has a causal interpretation, functional realization, and probabilistic consequences.
If $Z$ is the sole parent of $X$ and $Y$, then (i) $Z$ represents a generative process whose outcome causes the outcomes of $X$ and $Y$, (ii) the evaluation of $X(\omega)$ (or $Y(\omega)$) requires the evaluation of $Z(\omega)$, but not vice-versa, and (iii) $X$ and $Y$ are conditionally independent given $Z$.
A particularly common relationship pattern is a class of random variables that are mutually conditionally independent and identically distributed (c.i.i.d) given their parents.
This arises when an experiment is repeated (e.g. a die is tossed several times), or multiple noisy observations of the same process are taken. 
As such, unique elements of a c.i.i.d. classes are the primitive representation of probability distributions in  \omegalang{}.

\begin{definition} The c.i.i.d. class of $X$ given $Z$ is denoted $\ciidset{X}{Z}$ and inductively defined by:  
\begin{enumerate}
\item $X \in \ciidset{X}{Z}$
\item $A \in \ciidset{X}{Z}$ if for all $B \in \ciidset{X}{Z}$ where $A \neq B$:

(i) $A$ is independent of $B$ given $Z$

(ii) $A$ is identically distributed to $B$ given $Z$
\end{enumerate}
\end{definition}

To represent a random variable as a unique member of a c.i.i.d. class, we first observe that two random variables are c.i.i.d. given a third if they apply the same transformation to disjoint component of the sample space.  For example, $X(\omega) = \omega_1 + Z(\omega)$ and $Y(\omega) = \omega_2 + Z(\omega)$ are c.i.i.d. given $Z(\omega) = \omega_3$ since if Z is fixed (i) they map from disjoint components of $\Omega$ (they are conditionally independent), and (ii) the mapping is functionally identical (they are identically distributed).

Rather than \emph{syntactically} express that two random variables map from disjoint components of $\Omega$, we enforce it semantically by assigning each random variable $X$ a unique set of the dimensions of $\Omega$.
That is, we define $X$ as a function of a \emph{projection} of $\Omega$ -- characterized by an index set $I_X$ -- from a collection of unique projections: $((\Omega_1 \times \Omega_2, \times \cdots), (\Omega_j \times \Omega_{j+1} \times \cdots), ...)$.
For $X$ to be a defined on a projection means that any application $X(\omega)$ is replaced with $X(\omega_{I_X})$ where $\omega_{I_X} = \textrm{proj}_{I_X}(\omega) = (\omega_i, \omega_j, \omega_k,...)$ for all $i,j,k,... \in I_X$.

For the same reasons as above, two random variables will then be i.i.d. if they apply the same transformation to different projections of $\Omega$.
For example, if $f(\omega) = \omega_1 + \omega_2$, then $X(\omega) = f(\textrm{proj}_{I_X}(\omega))$ and $Y(\omega) = f(\textrm{proj}_{I_Y}(\omega))$, are i.i.d. if $I_X$, $I_Y$ are disjoint sets.

Projection is complicated by the fact that random variables typically have parents.
If $Z$ is the parent of $X$, e.g. in $X(\omega) = \omega_1 + Z(\omega)$, then the evaluation of $X(\omega)$ necessitates the evaluation of $Z(\omega)$.
Both $X$ and $Z$ are defined on their respective projections $I_X$ and $I_Z$, but if we first project $\omega$ onto $I_X$ to evaluate $X(\omega)$, we discard the elements necessary to project onto $I_Z$ in the evaluation $Z(\omega)$ within $X$.

To resolve this, we represent an element of the sample space as a reversible projection $\omega^\pi = (\omega, I)$, where $\omega \in \Omega$ is an element of the unprojected sample space, and $I$ is an index set.
$\omega^\pi_i = (\textrm{proj}_I(\omega))_i$ is the ith element of the projection onto $I$.
Crucially, $\omega^\pi$ can be projected from $I$ onto another index set $J$ trivially by substituting $I$ with $J$.


We represent, a random variable $X$ as  $(f_X, I_X)$, where $f_X:\Omega \to \tau$ is a function and $I_X $ is an index set of natural numbers.
$X$ is defined on a (potentially) concentrated probability space denoted $\mathcal{S}(X) = \SU \cap \{\omega \mid  X(\omega) \neq \bot, \omega \in \Omega\}$.
Random variable application reprojects $\omega$ onto $I_X$:
$$
X(\omega) := f_X(\textrm{proj}_{I_X}(\omega))
$$

\subsubsection*{Denotations}
The operator $\rv$ pairs a function of $\Omega$ (constructed using lambda abstraction) with a probability space to yield a random variable.
For example: $\sem{\rv(\lambda \omega : \Omega . 0)}$ denotes the constant random variable $(\omega \mapsto 0, I)$.

For a random variable to not be constant, it must access the sample space.
$\sem{\omega(i)}$ (rule $\omega app$) denotes an element in $[0, 1]$ of the sample space.
In particular, it is the $i$th element of a projection of $\Omega$.
By rule rvapp, this is the  projection of the random variable that the application $\omega(i)$ is within.
Since different random variables are defined on different projections, $\omega(i)$ within the context of distinct random variables are independent.
The value of $i$ is arbitrary, but indices must not be reused if independent values are needed.
The macro $\uuid$ resolves to a globally unique integer index, which relieves the programmer from specifying arbitrary indices.
For example, $\sem{\rv(\lambda \omega : \Omega . \omega(\uuid) + \omega(\uuid))}$ denotes the random variable $(\omega \mapsto \omega_1 + \omega_2, I)$.

Random variables can be treated like values from their realization space using pointwise random variable (rule pointwise) composition, e.g.: $\sem{\textrm{sqrt}(\rv(\lambda \omega : \Omega . \omega(\uuid))}$ denotes the random variable $(\omega \mapsto \sqrt{\omega_1}, I)$.

$\sem{\ciid(t)}$ is functionally identical to $\sem{t}$ but on a different projection of $\Omega$, and hence c.i.i.d. given the parents of $\sem{t}$.
For example, if $\stdunif = \lambda \omega : \Omega . \omega(\uuid)$, then $\sem{\stdunif + \stdunif}$ is a equivalent to $\sem{2 * \stdunif}$, whereas $\sem{\ciid(\stdunif) + \ciid(\stdunif)}$ is a sum of independent uniforms, a triangular distribution.


\textrm{rand} samples from a distribution.  In rule (rand), $\omega \sim \mathcal{S}'$ denotes that $\omega$ is an i.i.d. sample drawn from the probability space $\mathcal{S}'$.
That is, we assume the existence of an oracle able to randomly select $\omega \in \Omega'$ with probability $\mathcal{P}'(\{\omega\})$.
For example $\sem{\textrm{rand}(\rv(\lambda \omega : \Omega . \omega(\uuid) > 0.4))}$ denotes either true or false with probabilities 0.6 and 0.4 respectively.

$\sem{t_1 \mid t_2}$ denotes a random variable $\sem{t_1}$ conditioned on $\sem{t_2}$ being true by concentrating the probability space of $\sem{t_1}$ onto the event indicated by $\sem{t_2}$ (rule cond).
It is realized by a built-in function, which returns a random variable which maps $\omega$ to $\bot$ if $\sem{t_2}(\omega) = 0$.
For example, if $X$ is a normally distributed random variable then $\sem{X \mid X > 0}$ is a truncated normal defined on a concentrated space $\mathcal{S}'$.

The random conditional distribution operator $\parallel$ is defined in terms of $\mid$ and lambda abstraction as a built-in function (figure \ref{macros}).

$\sem{\dist(t)}$ calculates the distributional property of random variable $t$.
There are two cases depending on the type of $t$.
When $t$ is a regular random variable whose realization space is primitive types,
we rely on the underlying inference engine (denoted by $\sem{\dist}$) to return
the desired property (rule ($\dist$)).
For the specific engine we use, it approximates these properties by sampling.
When $t$ is a higher-order random variable whose realization space is random variables,
rule pointwise would be triggered if $\dist$ was a regular function.
However, it is not so rule (higher-order dist) is defined to provide for a similar effect.
It pushes $\dist$ inside the definition of $t$ recursively until it reaches the inner-most random variable.
This effectively creates distributions of distributional properties.

We now show that a random variable defined in \omegalang{} represents a well-defined probability distribution.
To simplify the discussion, without reducing the expressiveness, we do not allow $\rand$ to be invoked inside a function $f$ that is used to
construct a random variable ($\rv(f)$).
We further limit the discussion to function $f$ that always terminates.

\begin{definition}[Well-formed randvars] We say a random variable $t = \rv(f)$ is well-formed, if for any $\omega \in \Omega$, $f(\omega)$ terminates and $\rand$ is not invoked when evaluating $f(\omega)$.
\end{definition}

\begin{theorem}[Well-definedness]
	The distribution represented by a well-formed random variable is well-defined.
\end{theorem}
\begin{proof}
	Assume the variable is $t \mapsto (f,I)$.
	We prove the theorem by showing the probability density function/probability mass function of $t$ is well-defined.
	For simplicity, we assume $t$ is continuous and the proof for the case where $t$ is discrete is similar. 
	According to ($\rand$) rule, $\rand$ will only draw $\omega$ that falls into the domain of $f$:
	\[\omega \sim \mathcal{S}((f, I)) = \SU \cap \{\omega \mid  (f, I)(\omega) \neq \bot, \omega \in \Omega\}.\]
	Then the range of $\rand(t)$ is $R = \{ (f, I)(\omega) \mid \omega \in \Omega \wedge  (f, I)(\omega) \neq \bot\}$.
	Let $pdf$ be the pdf of $t$
	and $V = \Bool \cup \Int\linebreak \cup \Real \cup (\RV \tau)$.
	For $v \in V \setminus R$, we have $pdf(v) = 0$.
	For $v \in V \cap R$, its support is $D = \{\omega \mid  (f, I)(\omega) = v, \omega \in \Omega\}$.
	Since $t$ is well-formed, the evaluation of $(f,I)(\omega)$ is deterministic and always terminates. 
	Recall that $\SU$ is a uniform distribution on $d$-dimensional unit $\Omega$, then we have $pdf(v) = \sum_{\omega \in D} (\frac{1}{1-0})^d$.
	Thus, $pdf$ is well-defined over $V$.
\end{proof}


\begin{figure}[t]
	{\small
    \begin{align*}
        \sem{\rv(t)} &\mapsto (\sem{t}, I)  \text{ where } I \text{ is unique } \tag{rv}\\
        \sem{\ciid(t)} &\mapsto (f, J) \tag{ciid} 
        \\ &\text{ where } \sem{t} = (f, I) \text{ and } J \text{ is unique } \\
        \sem{t_1(t_2)} &\mapsto (\textrm{proj}_I(\omega))_i \tag{$\omega$app} \\
        &\text{ if } \sem{t_1} \mapsto (\omega, I) \text{ and } \sem{t_2} \mapsto i \\
        \sem{t_1(t_2)} &\mapsto \sem{e / (\omega, I))} \text{ if }\sem{t_1} \mapsto (\lambda x:\Omega.e, I)  \\
        & \text{ and } \sem{t_2} \mapsto (\omega, J) \tag{rvapp}\\
        \sem{t_1(t_2)} &\mapsto (\lambda \omega:\Omega.g(f_X(\omega)), I) \tag{pointwise} \\
            & \text{ if } \sem{t_1} \mapsto g:\tau_1 \mapsto \tau_2\\
            & \text{ and } \sem{t_2} \mapsto (f_X:\Omega \mapsto \tau_1, I) \\
        \sem{\textrm{rand}(t)} &\mapsto \sem{f((\omega, N))} \text{ where } \omega \sim \mathcal{S}((f, I)) \tag{rand} \\
            & \text{if } t \mapsto (f, I)  \\
        \sem{\dist(t)} & \mapsto \sem{\dist}((f,I)) \text{ if } \sem{t} \mapsto (f:\Omega \mapsto T, I)    \tag{dist}\\
        & \text{ where } T \text{ is } \Bool, \Real, \text{ or } \Int\\
        \sem{\dist(t)} & \mapsto (\lambda \omega:\Omega.\sem{\dist(f(\omega))}, I)  \tag{higher-order dist}\\
        & \text{ if } \sem{t} \mapsto (f:\Omega \mapsto \RV \tau, I)\\
    \end{align*}
}
\vspace{-0.5in}
    \caption{Denotational Semantics}
    \label{densemantics}
    \vspace{-0.2in}
    \end{figure}


\subsection{Example \omegalang{} Programs}
Here we demonstrate \omegalang{} through examples.

Sample from a standard uniform distribution:
{\small
\begin{omegacode}[numbers=left,xleftmargin=2em,firstnumber=last]
let x = rv(\_w:_W._w(uid)) in rand(x)
\end{omegacode}
}
\noindent
Distribution families map parameters to random variables:
{\small
\begin{omegacode}[numbers=left,xleftmargin=2em,firstnumber=last]
bern = \p:Real.rv(\_w:_W._w(uid) > p)
uniform = \a:Real, b:Real.rv(\_w:_W._w(uid)*(b-a)+a)
\end{omegacode}
}

Pass random variables as parameters to a family for Bayesian parameter estimation:
{\small
\begin{omegacode}[numbers=left,xleftmargin=2em,firstnumber=last]
let _m = uniform(0, 1)
  x1 = uniform(_m, 2)
  x2 = uniform(_m, 2) in
  rand(_m | rand(x1 = 1.3) && rand(x2 = 1.6))
\end{omegacode}
}

\noindent
Variance is expressible in terms of $\mathbb{E}$:
{\small
\begin{omegacode}[numbers=left,xleftmargin=2em,firstnumber=last]
var = \x:RV Real.E((x - E(x))^2)
\end{omegacode}
}

Laws of total expectation and variation.
Given  \texttt{x} and \texttt{y}:
\noindent
Variance is expressible in terms of $\mathbb{E}$:
{\small
\begin{omegacode}[numbers=left,xleftmargin=2em,firstnumber=last]
E(x) == E(E(x || y))
var(y) == E(var(y || x)) + var(E(y || x))
\end{omegacode}
}
\subsection{Inference}\label{inference}
Here we outline the general interface to inference in \omegalang{}

A conditional sampling algorithm in \omegalang{} is an implementation of the procedure $\rand$.

\begin{definition}A conditional sampling algorithm is function $\rand : (\Omega \to \tau) \to \tau$ which maps a random variable to a value sampled from its domain.
\end{definition}

\paragraph{Rejection Sampling}
$\rand_\textrm{rej}$ draws exact samples by rejecting those violating specified conditions.
It assumes a source of randomness  $\mathcal{A}$: an infinite matrix where $A_{i, j} \in [0, 1]$ is randomly and uniformly selected by an oracle.
A function $\rand_\omega(j) = \omega$ where $\omega_i = \mathcal{A}(i, j)$ maps integers to independent sample space elements.
$\rand_\textrm{rej}$ is defined recursively:
\begin{align*}
\rand_\textrm{rej}(x) &= \rand_\textrm{rej}(x, 1)\\
\rand_\textrm{rej}(x, j) &= \begin{cases} x(v) &\mbox{if } x(v) \neq \bot \\
\rand_\textrm{rej}(x, j + 1) & \mbox{otherwise }  \end{cases}\\ &\text{ where } v = x(\rand_\omega(j))\\
\end{align*}

\paragraph{Constraint Relaxation}
The expected time to draw a sample with rejection sampling is inversely proportional to the measure of the conditioning set, which can be vanishingly small.
Instead, most probabilistic programming languages apply more sophisticated inference methods to sample from posterior densities derived automatically from the model (e.g. by \cite{wingate2011lightweight}).  However, often this is inapplicable to distributional inference problems because the likelihood terms are intractable.  Likelihood-free inference procedures are then the sole option.

We use a recent likelihood-free approach given in \cite{tavares2018smooth}.
From a predicate $Y$ it constructs an energy function $U_Y: \Omega \to \mathbb{R}$ which approximates $Y$.
It relies on a distance metric, such that regions closer to the event indicated by $Y$ become more probable.
$U_Y$ is then sampled from using a variant of replica exchange Markov Chain Monte Carlo.  We defer to \cite{tavares2018smooth} for more details.

\section{Evaluation}\label{exp}
We first compare the expressiveness of \omegalang{} against an existing probabilistic language using a representative benchmark suite
both qualitatively and quantitatively.
Then we show that \omegalang{} is able to enable emerging applications using two case studies.

\begin{figure}[t]
\includegraphics[width=0.8\linewidth]{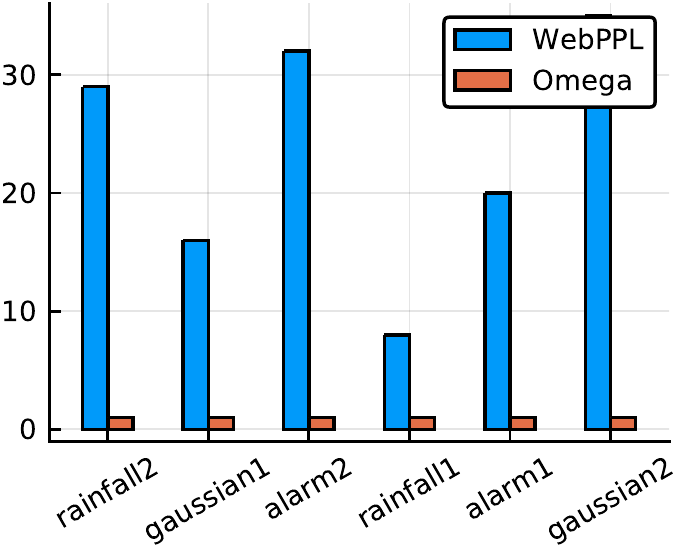}
\caption{Empirical comparison of lines of code}
\label{fig:loc}
\end{figure}

\subsection{Expressiveness}
Random conditional distributions increase the expressiveness of probabilistic programming.
Here we support this claim by comparison with WebPPL.
In particular, we show that while it is possible to express rcd queries in WebPPL by carefully
constructing the model, it requires
substantial effort to change the program to express queries that are different from the original ones.
In contrast, the effort to achieve similar effect in \omegalang{} is much less.

\subsubsection{Qualitative Analysis}
Consider an \omegalang{} program that samples from the conditional expectation $\ee(rain \mid winter)$:

{\small
\begin{omegacode}[numbers=left,xleftmargin=2em,firstnumber=last]
let bern = \p . \_w:_W._w(uid) > p
  winter = bern(0.5)
  clouds = if winter then bern(0.8) else bern(0.3)
  base = if winter then 3 else 0
  altitude = uniform([base + 3, base + 5, 10])
  rainfall = if clouds then altitude else 0 in
    mean(rainfall || clouds)
\end{omegacode}
}

Here, \texttt{mean} approximates the expectation of \texttt{rainfall} by sampling.
To sample a different distributional property, e.g. \texttt{E(rainfall || winter)} or \texttt{var(rainfall || clouds)},
we only need to change the final query expression without touching the original model.
Contrast this with the same example in WebPPL:

{\small
\begin{omegacode}[numbers=left,xleftmargin=2em,firstnumber=last]
var model = function() {
  winter = bernoulli(.5)
  rainfall = Infer(function(){
    clouds = winter ? bernoulli(0.8) : bernoulli(0.3)
    base = winter ? 3 : 0
    alt = uniformDraw([3 + base, 5 + base, 10])
    return clouds ? alt : 0.0
    })
  return expectation(rainfall)
}
\end{omegacode}
}

By excluding \texttt{winter} within the definition of \texttt{rainfall}, we have split the model into two.  A single sample from \texttt{model} will sample \texttt{winter} once, and with this value fixed will sample \texttt{clouds}, \texttt{altitude}, and \texttt{base}  several times to approximate the expectation.  This achieves the same effect as rcd.

The WebPPL model is ostensibly of similar complexity to the \omegalang{} model.
However, we have had to bake the conditional expectation into the generative model.
This violates the principle discussed in section \ref{higher} that distributional properties are derived from, not in addition to the original base model.
We explicitly constructed a model split into two parts whereas $\parallel$ can construct any such split automatically.
Of practical importance, changing the distributional property can require substantial changes to the model.

For example, to sample instead from $\ee(rain \mid clouds)$ in \omegalang{} is trivial, whereas in WebPPL the model changes dramatically:

{\small
\begin{omegacode}[numbers=left,xleftmargin=2em,firstnumber=last]
var winter_and_clouds = function()
{
  var winter = bernoulli(.5)
  var clouds = winter ? bernoulli(0.8) : bernoulli(0.3)
  return {winter: winter, clouds: clouds}
}
var conditioned_model = function(iscloudy)
{
  w_and_c = winter_and_clouds()
  winter = w_and_c.winter
  clouds = w_and_c.clouds
  condition(clouds == iscloudy)
  base = winter ? 3 : 0
  altitude = uniformDraw([3 + base, 5 + base, 10])
  return clouds ? altitude : 0.0
}
var model = function() {
  w_and_c = winter_and_clouds()
  winter = w_and_c.winter
  clouds = w_and_c.clouds
  rainfall = Infer(function(){
    return conditioned_model(clouds)
    })
  return expectation(rainfall)
}
\end{omegacode}
}

We must then manually construct a nested inference problem to sample the distributional property.  

\subsubsection{Quantitative Comparison}
To evaluate this quantitatively, we evaluated the degree to which several models must be modified to construct random distributional proprties.
For model $\mathcal{M}$ from a set of models (below), we:
\begin{enumerate}
\item Select two variables at random $\Theta, X \in \mathcal{M}$
\item From $\mathcal{M}$ manually construct $\mathcal{M}'$ where $\rcdxy{X}{\Theta} \in \mathcal{M}'$
\item Evaluate the number of lines from $\mathcal{M}$ to $\mathcal{M}'$
\end{enumerate}

Models:

\begin{itemize}
\item \textit{rainfall}: as above
\item \textit{alarm}: a model over alarm clock and time
\item \textit{gauss}: a binomial-gaussian model
\end{itemize}

Figure \ref{fig:loc} demonstrates expressiveness improvements in \omegalang{} vs WebPPL.
WebPPL requires a large number of edits for every change, whereas \omegalang{} requires exactly 1.



\subsection{Case Studies}\label{study}

This section explores two studies of distributional inference.
All inferences were performed on a Linux 4.15 laptop with a 1.9GHZ quad-core I7 processor and 24GB memory.

\subsubsection{Inferring Fair Classifiers}
In this case study, we aim to transfer an unfair
non-probabilistic classifier into a fair one.

\paragraph{Setup.}
The classifier is studied by FairSquare~\cite{fairsquare}, a tool for verifying
algorithmic fairness.
It is an SVM ($\text{SVM}_4$ in Section 6 of \cite{fairsquare}) that predicts whether a person has a high income based on their gender, age, capital gains, and capital losses.
The classifier is trained on a popular income dataset~\footnote{\url{https://archive.ics.uci.edu/ml/datasets/Adult}} and judged as unfair by FairSquare.
Same as the setting in \cite{fairsquare}, we use equality of opportunity~\cite{hardt2016} as the fairness specification:
\begin{equation}\label{fair1}
\frac{\prob(high\ income \mid female \wedge age > 18)}{\prob(high\ income \mid male \wedge age > 18)} > 0.85.
\end{equation}
%

\begin{figure}[t!]
{\small
\begin{omegacode}[numbers=left,xleftmargin=2em,firstnumber=last]
# input distribution
let gender, age, cap_gain, cap_loss = popModel()
    # parameter distribution
    rparams = [normal(p, 1.0) for p in params] in
	
# output distribution
let high_income = SVM(rparams, gender, age, 
                           cap_gain, cap_loss) in
	
# fairness specification
let f_high_income = cond(high_income, 
                    gender == female and age > 18)
    m_high_income = cond(high_income, 
                    gender == male and age > 18) in
                    
let fairness = (prob(qual_female_high_income || rparams) 
     / prob(qual_male_high_income || rparams)> 0.85) in
	
# parameter distribution conditioning on being fair
let fair_rparams = cond(rparams, fairness) in
	
# draw fair parameter samples
rand(fair_rparams)
\end{omegacode}
}
\caption{Probabilistic program for inferring fair classifiers.}
\label{prog:fair}
\end{figure}

\paragraph{Probabilistic Program.}Figure~\ref{prog:fair} outlines the \omegalang{} program which infers parameters which make the SVM fair.
Given fairness is a distributional property, we begin with constructing random variables which represent the input distribution by invoking \textsf{popModel}().
To implement \textsf{popModel}, we use a Bayesian network that is described in \cite{fairsquare}.
Then we construct a distribution of SVM parameters.
While our end goal is to make the SVM fair, we also do not want its accuracy to degrade much.
Therefore, we create an array of normals as the random parameters whose means are the original learned non-random parameters.
This will make our final samples of parameters similar to the original ones.
Next, we construct the fairness specification and lift it as a distribution which depends on the random parameters.
Here we use \inlinecode{prob} to evaluate the probability of whether a statement holds,
which is approximated by taking the mean of samples drawn from it.
Finally, we draw samples from the random parameters conditioning on the fairness specification.

\paragraph{Accuracy and efficiency.} To validate our approach, we use FairSquare~\cite{fairsquare} to verify whether the produced SVM is fair.
%
%
In theory, our result can be inaccurate due to two reasons: we approximate \texttt{prob} by samples, and the underlying solver itself does approximate inference.
However, in this application, we observe running our program produces high-quality results:
we sampled 10 parameters using each algorithm, and FairSquare returned fair for all of them.
Moreover, it only takes 30 seconds to produce all 10 samples.
While $\rcd$ improves the expressiveness over existing probabilistic constructs,
it can be still evaluated in a accurate and efficient manner.

%
%
%
%

%
%

\subsubsection{Inferring Robust Classifiers}
In this case study, we show how to improve the robustness of a classifier using \omegalang.
Machine learning models such as neural networks have been shown being vulnerable to adversarial inputs~\cite{szegedy2013intriguing}:
a small perturbation to the input can completely change the output.
For security concerns, it is desirable that a machine learning model is resilient to
such perturbations, in other words, robust.
We do not claim that we have solved the problem nor that we intend to compete with existing techniques
due to limitations in the underlying solvers.
Instead, we use this use case to demonstrate the expressiveness of our language.
Concretely, we will improve the robustness of a small neural network against a specific adversarial attack.

\paragraph{Setup.} The neural network has three layers and two hidden neurons.
Similar to the previous case study, it is also studied in \cite{fairsquare} (NN$_{3,2}$ in Section 6) and trained on the same income dataset.
We apply fast gradient sign method \cite{goodfellow2014explain} to generate adversarial inputs.
Out of 1,000 inputs that are randomly drawn from the input distribution, the network is resilient to the attack on 97\% of the inputs.
Our goal is to improve this ratio above 99\%.
We do not intend to make the network robust on arbitrary input because when the input space is continuous, inevitably there will be inputs that are very close to the decision boundary.

\begin{figure}[t!]
{\small
\begin{omegacode}[numbers=left,xleftmargin=2em,firstnumber=last]
# input distribution
let gender, age, num_edu, cap_gain = popModel()
    # parameter distribution
    rparams = [normal(p, 1.0) for p in params] in

# output distribution
let high_income = NN(rparams, gender, age, num_edu, 
                                     cap_gain) in

# craft adversarial attack
let perturb = fast_gradient_sign(rparams, gender, age, 
                                num_edu, cap_gain) in

let adv_gender, adv_age, adv_num_edu, adv_cap_gain = 
                 gender+perturb[1], age+perturb[2], 
            num_edu+perturb[3], cap_gain+perturb[4] in

# robustness specification
let adv_high_income = NN(rparams, adv_gender, adv_age, 
                         adv_num_edu, adv_cap_gain) in

let robustness = (prob((adv_high_income == high_income) 
                                || rparams) > 0.99) in

# parameter distribution conditioning on being robust
let robust_rparams = cond(rparams, robustness) in

# draw robust parameter samples
rand(robust_rparams)
\end{omegacode}
}
\caption{Probabilistic program for inferring robust classifiers.}
\label{prog:robust}
\end{figure}

\paragraph{Probabilistic Program.}Figure~\ref{prog:robust} outlines the \omegalang{} program we constructed for inferring parameters which make the neural network robust.
It begins with constructing random variables which represent the input distribution by invoking \textsf{popModel}().
Similar to the previous case study, \textsf{popModel} is implemented using a Bayesian network described in \cite{fairsquare}.
Then we construct distributions of neural network parameters, each of which is a normal whose mean is the corresponding original non-random parameter that is obtained by training on the dataset. 
Next, we construct perturbations to the inputs by using
fast gradient sign method~\cite{szegedy2013intriguing}.
These perturbations are random variables that depend on the inputs and the parameters.
Then we construct the robustness specification, which states that the output to the adversarial input should be the same as the original output with a probability greater than 0.99.
Finally, we draw parameter samples from the random parameters conditioning on the robustness specification.

\paragraph{Accuracy and efficiency.} To evaluate whether a sampled parameter leads to a robust classifier, we drew 1,000 samples from the input distribution, and applied fast gradient sign method to see if the attack changed the outputs.
We drew 10 parameters and observed the resulted networks to be robust on 96.6\%-100\% inputs, with 98.8\% being the average.
Due to the approximation in evaluating \texttt{prob} and the underlying solver, our approach
does not guarantee that the robustness specification is always satisfied.
However, it does improve the robustness of the network for most cases.
Moreover, these 10 parameters are produced in only 86 seconds. 
%

%
%

%
%
%
%


\section{Related Work}
This contribution builds upon a long history of incorporating probability into programming languages \cite{kozen1979semantics, goodman2008church, mansinghka2014venture,wood2014new}.
Most probabilistic programming languages define samplers: procedures which invoke a random source of entropy \cite{park2005probabilistic}.
In contrast, in \omegalang{} there is a strict separation between modeling and sampling.
Defining a probabilistic program means to construct a collection of random variables on a shared probability space.
A variety of both old and recent work \cite{kozen1979semantics,staton2016semantics} has gone into defining measure-theoretic based semantics for sampling based probabilistic programming languages.  \omegalang{} instead has measure-theoretic objects as its primitive constructs.

For particular cases, \emph{nested inference} has been used to achieve the same effects as random conditional distributions.
They emerged out of the need to express nested goals \cite{mantadelis2011nesting} for ProbLog \cite{kimmig2008efficient} and to model recursive reasoning \cite{stuhlmuller2014reasoning}.
The work of \cite{mantadelis2011nesting} aimed to mimic the capability of Prolog in handling nested goals: a outer goal could use the inner goal's success probability.
To this end, they allow calls to problog inference engine within the inside a model.
In a similar way, \cite{stuhlmuller2014reasoning} added nested capabilities to Church by first defining \inlinecode{query} as a Church function.  More recently, the programming language WebPPL \cite{dippl} supports inference inside a model, which allows to compute the Kullback-Leibler divergence between two random distributions for optimal experiment design \cite{ouyang2016practical}. A review on \textit{nested probabilistic programs} with focus on inference can be found in \cite{rainforth2018nesting}


Algorithmic fairness and robustness have received much attention recently.
In machine learning, various fairness specifications~\cite{calders2009,hardt2016,kilbertus2017,kusner17} have been proposed and many methods~\cite{calders10,davies17,dwork2012,dwork2018,fish16} have been designed to train a model to satisfy these specifications.
These methods typically formulate the problem of training a fair classifier as a constrained or unconstrained optimization problem over the training data set to balance the tradeoff between accuracy and fairness.
Similarly, many techniques have been proposed to improve robustness of machine learning models, in particular, neural networks.
These techniques include obfuscating gradients~\cite{athalye2018}, adversarial training~\cite{szegedy2013intriguing,goodfellow2014explain},
cascade adversarial training~\cite{na2017},
constrained optimization~\cite{raghunathan2018}, and many others.
We do not intend to compete against these methods for improving fairness and robustness, but instead use these two case studies to demonstrate the expressiveness of our approach.
However, our approach does have an edge over existing approaches when the user has prior knowledge about the input distribution.

In some cases it is possible to manually encode declarative knowledge constructively into a generative model.
For example, one may specify a truncated normal distribution by conditioning a normal distribution to be bounded, or constructively using the inverse transform method.
However, in most cases there is no straightforward means.
Continuing the glucose example, we may attempt to tie parameters by drawing the parameters for each patient from a shared stochastic process.
However, this increases the complexity of the generative model and requires significant expertise to ensure that the new generative model indeed captures the high-level constraint without unnecessarily constraining the model in unwanted ways.



\bibliography{bib}



\end{document}


\title[The Random Conditional Distribution]{The Random Conditional Distribution}        
\subtitle{For Uncertainty in Distributional Properties \linebreak (Appendix)}                     


\author{Zenna Tavares}
\affiliation{
  \department{CSAIL}              
  \institution{MIT}            
}
\email{zenna@mit.edu}          

\author{Xin Zhang}
\affiliation{
  \department{CSAIL}              
  \institution{MIT}            
}

\author{Edgar Minaysan}
\affiliation{
  \department{CSAIL}              
  \institution{MIT}            
}

\author{Javier Burroni}
\affiliation{
  \department{CSAIL}              
  \institution{MIT}            
}

\author{Rajesh Ranganath}
\affiliation{
  \department{Courant Institute}              
  \institution{NYU}            
}

\author{Armando Solar Lezama}
\affiliation{
  \department{CSAIL}              
  \institution{MIT}            
}


\maketitle
\setmonofont[Scale=MatchLowercase]{DejaVu Sans Mono}

\appendix
\section{Operational Semantics}\label{operationalsemantics}

Big step operational semantics of \omegalang{} are defined in figure \ref{semantics}.
Operational semantics of standard terms are standard and omitted.

\begin{figure}[h]
\begin{gather*}
    \infer[rv]
    {\eto{\rv(\lambda x : \tau. t)} {\crvv{\lambda x:\tau.t} {newid()} }}
    {}\\
    \infer[ciid]
    {\eto{\ciid(t_1)} {\crvv{v} {newid()}}}
    {\eto{t_1}{\crvv{v} {i} }}
    \\
    \infer[cond]
    {\eto{t_1 \mid t_2}{\crvv{\lambda \omega: \Omega. \text{ if } y(\omega) \text{ then } x(\omega) \text{ else } \bot}{newid()}}}
    {\eto{t_1}{x} & \eto{t_2}{y}}
    \\
    \infer[\rv app]
    {\eto{(t_1 \; t_2)} {v}}
    {\eto{t_1}{\crvv{\lambda \omega: \Omega . t} {i} } & \eto{t_2} {\crvv{f}{j}} & \eto{\sub{\omega}{\crvv{f}{i}}{t}}{v}}
    \\
    \infer[\omega app]
    {\eto{(t_1 \; t_2)} {\omega(i, j)}}
    {\eto{t_1}{(\omega, i)}  & \eto{t_2}{j : \Int}}
    \\
    \infer[pointwise]{\eto{(t_1 \; t_2)}{\crvv{\lambda \omega: \Omega . f(X(\omega))}{j}}}
    {\eto{t_1}{\crvv{f: \tau_1 \to \tau_2}{i}} & \eto{t_2}{\crvv{X: \Omega \to \tau_1}{j}}}
    \\
    \infer[rand]{\eto{rand(t)}{v}} 
                {t(x) \mapsto v} \\
    \text{where } x = \crvv{\lambda i: \Int . \lambda j: \Int. \mathcal{A}(i, j, k) } {k} \text{ and } \\ k \text{ is uniformly drawn from } \{ k \mid k \in N \wedge  v \neq \bot\}   
                \\
                \\
                \infer[\bot app_1]{\eto{(t_1 \; t_2)}{\bot}}
                                {\eto{t_1}{\bot}} 
                \qquad
                \infer[\bot app_2]{\eto{(t_1 \; t_2)}{\bot}}
                {\eto{t_2}{\bot}}\\
                \infer[\bot ite_1]{\eto{\ite{t_1} {t_2} {t_3}}{\bot}}
                {\eto{t_1}{\bot}}\\
                \infer[\bot ite_2]{\eto{\ite{t_1} {t_2} {t_3}}{\bot}}
                {\eto{t_1}{\true} & \eto{t_2}{\bot}}\\
                \infer[\bot ite_3]{\eto{\ite{t_1} {t_2} {t_3}}{\bot}}
                {\eto{t_1}{\false} & \eto{t_3}{\bot}}
\end{gather*}
    \caption{Operational Semantics for \omegalang{}}
    \label{semantics}
\end{figure}


\subsection{Domains}

If the dimensionality of the model is infinite, we cannot explicitly represent the index set $I$ of a random variable neither, nor compute the projection mapping 
$\textrm{proj}_I$.
Instead we (i) index $\Omega$ with two indices: $\Omega = \Omega_{1, 1} \times \Omega_{1, 2} \times \cdots \Omega_{2, 1} \times \cdots$, and (ii)
encode a random variable as an  \emph{indexed function} $\crvv{f}{i}$ where $f$ is a function and $i$ is a unique identifier ($\id$).
The subspace of $\Omega$ allocated to $X = \crvv{f}{i}$ is then simply $\Omega_{i, 1} \times \Omega_{i, 2} \cdots$.

A concrete representation of a projection of $\Omega$ is then a pair $(\omega, i)$, where $i$ is the index of the random variable
Projection itself.
\subsection{Reductions}

Rule $(rv)$ constructs a random variable of type $\RV \tau$ by assigning a function $f$ a unique identifier $i$ to yield an indexed function $\crvv{f}{i}$.
The index is required to project the sample space down to the subspace of the random variable.
This occurs using non-standard semantics for application of random variables to values with rule (rv $app$).
Prior to substitution, application of random variable $X$ projects $\omega$ onto a subspace unique to $X$ by attaching its id to $\omega$ such that the index $i$ in application $\omega(i)$ within $X$ is relative to that subspace (specify by rule ($\omega app$)).
Crucially, $\omega(i)$ within $X$ is distinct from within the context of any other random variable.


By combining uniqueness of random variables ($rv$) and projection of sample space when indexing $\omega$ (rv$app$ and $\omega$app), we can create random variables that are c.i.i.d. (conditionally independent and identically distributed) by defining them in the same syntactic form.
We can achieve the same effect with $\ciid$, which simply copies a function and changes its id (rule ciid).
This allows us to easily construct random i.i.d. random variables.

\omegalang{} uses $\bot$ to signal errors. In particular, a partial function applied to an input on which it is not defined yields $\bot$.
\omegalang{} treats conditional random variables as partial functions defined only on the subset of $\Omega$ where the condition is satisfied.
Conditioning a random variable $X$ onto $Y$ simply returns a random variable with maps $\omega$ to $\bot$ if $Y(\omega) = \false$.
Rules for $(\bot app_1)$, $(\bot app_2)$, $(\bot ite_1)$, $(\bot ite_2)$, $(\bot ite_3)$ define operations on undefined values to always yield undefined values.
In this way, $\bot$ propagates, such that if a random variable $X$ depends on a conditioned random variable $Y$, $Y(\omega) = \bot$ implies that $X(\omega) = \bot$.


\omegalang{} separates sampling from the modeling.
Similar to denotational semantics, without committing to any specific algorithm we give a declarative definition of sampling using rule $rand$. 
Random variables are pure functions.
Hence, to draw a sample from a random variable we apply it to random input $\omega$ drawn according to its probability measure.
The operational semantics assumes a global infinite array $\mathcal{A}$ of random numbers in $[0, 1]$.
For convenience, $\mathcal{A}$ is three dimensional and $\mathcal{A}(i, j, k)$ denotes the i, j, kth element in that array.
Intuitively, $i$ corresponds to the id of the random variable that $\omega$ is passed to and $j$ is the index written in the program (i.e., $\omega{i}$).
As for $k$, it is used to generate a different $\omega$.
To respect the condition statements, the underlying inference algorithm should draw a $k$ that does cause the program evaluates to $\bot$. 



